\documentstyle[12pt,aaspp4_rj,psfig]{article}

\def\cge      {{$_ >\atop{^\sim}$}}
\def\cle      {{$_ <\atop{^\sim}$}}

\def\wfpc     {{\sl WFPC2}}
\def\etal     {{\it et\thinspace al.} }

\def\eg       {{\it e.g.}}
\def\err#1#2 {{$_{#1}\atop{^{#2}}$}}

\begin{document}

\title{The Ultraviolet Luminosity Density of the Universe from Photometric
Redshifts of Galaxies in the Hubble Deep Field\footnote{Based on observations
with the NASA/ESA {\it Hubble Space Telescope} obtained at the Space Telescope
Science Institute, which is operated by AURA, Inc., under NASA Contract NAS
5-26555.}$^,$\footnote{Based in part on observations with the KPNO 4-meter Mayall
Telescope of NOAO, which is operated by AURA,
Inc., under cooperative agreement with the NSF.}}

\author{Sebastian M. Pascarelle and Kenneth M. Lanzetta}
\affil{Department of Physics and Astronomy, State University of New York at
Stony Brook, Stony Brook, NY 11794-3800}
\authoremail{smp@mail.ess.sunysb.edu, lanzetta@mail.ess.sunysb.edu}

\and

\author{Alberto Fern\'andez-Soto}
\affil{School of Physics, University of New South Wales, Sydney, NSW2052,
Australia}
\authoremail{fsoto@bat.phys.unsw.edu.au}

\newpage

\begin{abstract}
Studies of the Hubble Deep Field (HDF) and other deep surveys have revealed an
apparent peak in the ultraviolet (UV) luminosity density, and therefore the
star-formation rate density, of the Universe at redshifts $1<z<2$. We use
photometric redshifts of galaxies in the HDF to determine the comoving UV
luminosity density and find that, when errors (in particular, sampling
error) are properly accounted for, a flat distribution is statistically
indistinguishable from a distribution peaked at $z\simeq 1.5$.
Furthermore, we examine the effects of cosmological surface brightness (SB)
dimming on these measurements by applying a uniform SB cut to all galaxy
fluxes after correcting them to redshift $z=5$. We find that, comparing all
galaxies {\em at the same intrinsic surface brightness sensitivity}, the UV
luminosity density contributed by high intrinsic SB regions increases by
almost two orders of magnitude from $z\simeq 0$ to $z\simeq 5$.
This suggests that there exists a population of objects
with very high star formation rates at high redshifts that apparently do
not exist at low redshifts. The peak of star formation, then, likely occurs
somewhere beyond $z>2$.
\end{abstract}


\newpage

\section{Introduction}

Important properties of the Universe as a whole can be determined by analyzing
complete samples of galaxy redshifts to very faint magnitude limits.  The
comoving ultraviolet (UV) luminosity density is one such property.  Because it
is dominated by massive, short-lived stars, the UV emission of an actively
star forming galaxy is nearly independent of star formation history.  For this
reason, the comoving UV luminosity density is directly related to the comoving
star formation and metal production rate densities of the Universe (\eg, Cowie
1988; Fall, Charlot, \& Pei 1996; Madau \etal 1996, hereafter M96). Lilly \etal
(1996, hereafter L96) measured the UV
luminosity density at redshifts $z$\cle 1 using the Canada--France Redshift
Survey (CFRS) and found that it rises rapidly by a factor of $\sim 15$ from
$z=0$ to $z=1$.  Similar results were found in this redshift range by Cowie,
Hu, \& Songaila (1997).

 M96 (later updated by Madau, Pozzetti, \& Dickinson 1998, hereafter M98)
measured the UV luminosity density at redshifts $z \simeq 2.75$ and
$z \simeq 4.00$ using the $U$- and $B$-band ``dropout'' technique applied to
the Hubble Deep Field (HDF) and found that it might be lower than the L96
measurements at $z$\cle 1 by a factor of $\sim 2$.
This suggested that there might be a peak in the UV luminosity density---and
therefore the star formation and metal production rate densities---at some
redshift between $z = 1$ and 2, prompting a spate of theoretical and
observational interpretation
(e.g., Shaver \etal 1996; Baugh \etal 1998; Ferguson \& Babul 1998; M98;
Madau, Della Valle, \& Panagia 1998; Silk \& Rees
1998). More recently, Connolly \etal (1997, hereafter C97) measured the
UV luminosity density from the HDF in the previously unsurveyed redshift range
1\cle $z$ \cle 2 and found that it appears to peak at $z \simeq 1.5$, which is
consistent
with the results of L96 and M98. However the photometric redshifts of galaxies
in the HDF from Sawicki, Lin, and Yee (1997) result in a UV luminosity density
that continues to increase to $z$\cge 2.5, indicating that the peak at $z
\simeq 1.5$ may be questionable.

In this paper, we apply our photometric redshifts of galaxies identified in
the HDF (Lanzetta, Yahil, \& Fern\'andez-Soto 1996, hereafter LYF96; Lanzetta,
Fern\'andez-Soto, \& Yahil 1998; Fern\'andez-Soto, Lanzetta, \& Yahil 1998,
hereafter FLY98) to measure the UV luminosity density of the Universe at
redshifts $0 < z < 6$.  Our analysis differs from previous analyses in three
important ways. First, our photometric redshifts are determined from spectral
template fits to optimal photometry of optical (Williams \etal 1996) and
infrared (Dickinson \etal 1998) images of the HDF.  In contrast to the $U$-
and $B$-band ``dropout'' technique of M98, our analysis determines the most
likely redshift of {\it every} galaxy in the HDF, and in contrast to the
photometric redshifts of C97, our analysis makes use of the $J$-, $H$-, and
$K$-band infrared photometry (instead of only the $J$-band photometry).
Second, we determine realistic uncertainties of the luminosity density
measurements, including the effects of systematic and photometric error on the
photometric redshifts and of sampling error.  Previous analyses have neglected
sampling error, which in fact dominates the uncertainty of the luminosity
density measurements in the HDF.  Third, we explicitly consider the effects of
cosmological $(1+z)^3$ surface brightness dimming on the observed luminosity
density.  Previous analyses have neglected cosmological $(1+z)^3$ surface
brightness dimming, which varies by more than two orders of magnitude over the
redshift range of galaxies identified in the HDF image.

In \S\ 2 we briefly describe our photometric redshift technique. In \S\ 3 we
present the results of our measurement of the UV luminosity density, in \S\ 4
we show how this determination is affected by cosmological surface brightness
dimming, and in \S\ 5 we discuss these results and present our conclusions.

\section{Photometric Redshifts}

  The starting point of our analysis is the photometric redshift estimates of
LYF96 and FLY98. Because details of the photometric redshift estimation
technique have been and will be presented elsewhere (Lanzetta,
Fern\'andez-Soto, \& Yahil 1998; FLY98), we simply summarize the method here.

  Galaxy photometry is determined from the optical F300W, F450W, F606W, and
F814W (Williams \etal 1996) and infrared $J$, $H$, and $K$ (Dickinson \etal
1998) images of the HDF.  To measure fluxes and flux
uncertainties in the optical images, we directly integrate within the aperture
mask of every object detected in the F814W image.  To measure fluxes and flux
uncertainties in the infrared images, we (1) model the spatial profile of every
object detected in the F814W image as a convolution of the portion of the F814W
image containing the object with the appropriate point spread function of the
infrared image and (2) determine a least-squares fit of a linear sum of the
model spatial profiles to the infrared image.  The advantages of this method
over simple aperture photometry are that the flux measurements correctly
weight signal-to-noise ratio variations within the spatial profiles, and the
flux uncertainty measurements correctly include the contributions of nearby,
overlapping neighbors.

  Galaxy redshifts are determined from fits to the spectral templates of E/S0,
Sbc, Scd, and Irr galaxies,
including the effects of intrinsic and intervening
neutral hydrogen absorption.  These effects are included as a function of
redshift, with mean values taken from Madau (1995) and Webb (1997). First, we
integrate the redshifted spectral
templates with the throughputs of the F300W, F450W, F606W, F814W, $J$, $H$, and
$K$ filters, at redshifts spanning $z = 0 - 7$.   Next, we construct the
``redshift likelihood function'' of every object detected in the F814W image by
calculating the relative likelihood of obtaining the measured fluxes and
uncertainties given the modeled fluxes at an assumed redshift, maximizing with
respect to galaxy spectral type and arbitrary flux normalization.  Finally, we
determine the maximum-likelihood redshift estimate of every object detected in
the F814W image by maximizing the redshift likelihood function with respect to
redshift. The result of the most recent application of this method is
presented in the catalog of FLY98, which lists photometric redshift estimates
of 1067 galaxies to a limiting magnitude threshold of $AB(814) = 28.0$ over an
angular area of $\sim$4 arcmin$^2$.

  Spectroscopic redshifts of more than 100 galaxies in the HDF have been
obtained using the Keck telescope (Cohen \etal 1996; Cowie 1997; Steidel \etal
1996; Zepf \etal 1997; Lowenthal \etal 1997).  Comparison between the
spectroscopic and photometric redshifts indicates the following results: (1)
At redshifts $z <  2$, the residuals between the spectroscopic and photometric
redshifts are characterized by an RMS dispersion of $\sigma = 0.09$ and a
discordant fraction ($>3\sigma$ discrepant after sigma clipping) of 0\%; (2)
at redshifts $z > 2$, the residuals between
the spectroscopic and photometric redshifts are characterized by an RMS
dispersion of $\sigma = 0.45$ and a discordant fraction of 7\%;  and (3)
these residuals between the spectroscopic and photometric redshifts arise from
cosmic variance with respect to the spectral templates (rather than from
photometric error), so a proper assessment of the errors of the photometric
redshifts of faint galaxies must include the effects of systematic and
photometric error.

\section{Ultraviolet Luminosity Density}

  We determined the luminosity (per unit wavelength interval) of each
galaxy at a rest-frame wavelength $\sim$1500\AA\ by applying an empirical
$K$-correction derived from the best-fit spectral template to the measured
galaxy photometry.  The $K$-corrections are {\em interpolated} from the
measured photometry at redshifts $z > 0.6$, although they are extrapolated from
the measured photometry (by up to a factor of two in wavelength) at redshifts
$z < 0.6$.  Next, we determined the comoving luminosity density versus redshift
by arranging the galaxies into redshift bins, summing the luminosities within
the bins, and dividing by the appropriate comoving volumes.  Finally, we
determined the uncertainty of the comoving luminosity density versus redshift
by applying a bootstrap resampling technique.  For each iteration of the
bootstrap technique, we resampled the photometric catalog and redetermined the
photometric redshift of each resampled galaxy, perturbing the photometry by
random deviates according to the measured photometric error and perturbing the
redshift by a random deviate according to the RMS residuals described in \S 2.
We then determined the comoving luminosity density versus redshift using the
resampled, perturbed redshift catalog.  We repeated the procedure 1000 times
in order to determine the range of values compatible with the observations.
This procedure explicitly allows for sampling error, photometric error, and
cosmic variance with respect to the spectral templates.

\begin{figure}
\centerline{\psfig{file=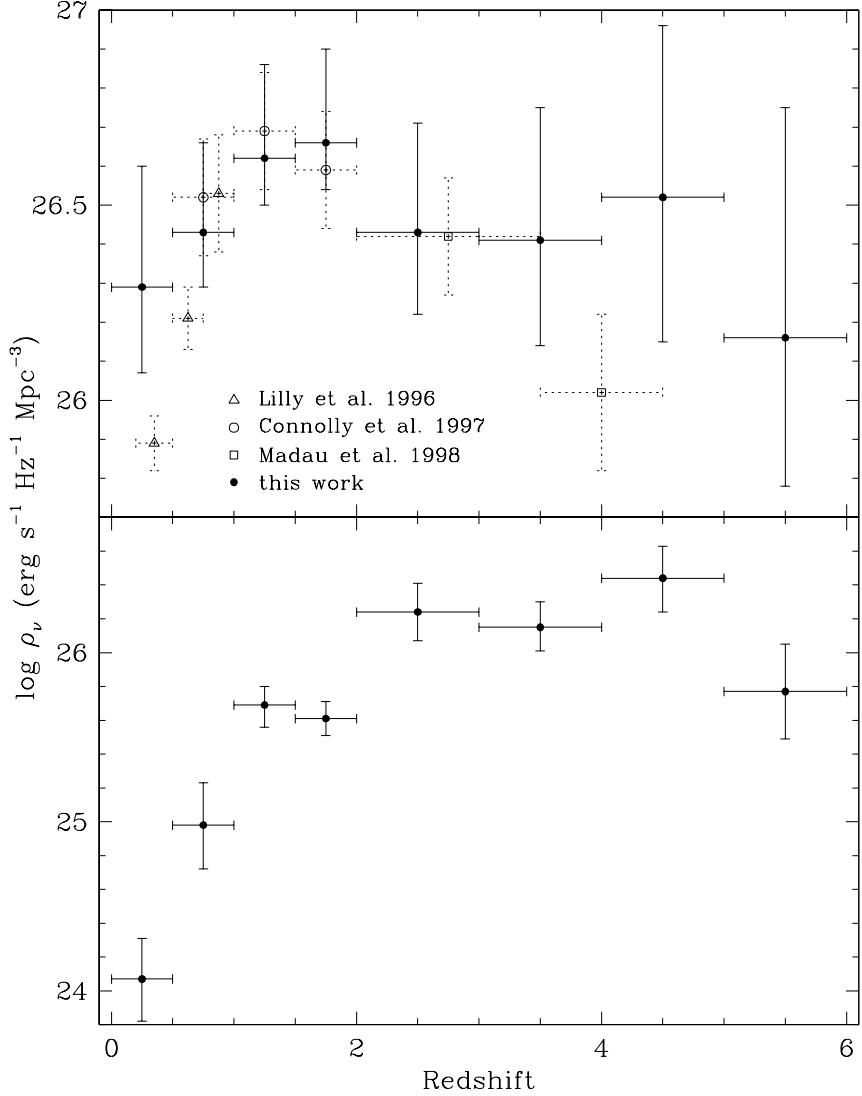,width=12cm}}
\caption{({\it a}), The UV luminosity density as a function of redshift as
measured from galaxies in the HDF using our photometric redshifts (filled
circles). The errorbars were calculated from a bootstrapping code which takes
into account the effects of systematic and photometric error on the
photometric redshifts and the effects of sampling error. For comparison, we
have included data points from Lilly \etal (1996, open triangles), Connolly
\etal (1997, open circles), and Madau \etal (1998, open squares). Note that
while our data appear to show a possible increase in the UV luminosity density
from $z=0-2$, there is little evidence for a decrease at higher redshifts to
within the errors. ({\it b}), The UV luminosity density arising from
intrinsically high surface brightness (SB) regions, after applying a uniform
SB cut so that $z<5$ galaxies are considered down to the same SB level as
$z$\cge 5 galaxies. Comparing the HDF galaxies at the same intrinsic SB
sensitivity shows that the UV luminosity density of high SB regions increases
by almost two orders of magnitude from $z\simeq 0$ to $z\simeq 5$.}
\end{figure}

  The results are shown in the top panel of Figure 1, which plots the comoving
UV luminosity density versus redshift, and are given in the second column of
Table 1. For comparison, the data points from L96, C97, and M98 are also
plotted in Figure 1{\it a}. Note that our data are entirely consistent with the
previous measurements to well within $\simeq 1.5\sigma$ (and all but our first
data point are within \cle1$\sigma$ of previous data). The fact that we do
not appear to reproduce the steep rise in the UV luminosity density from
$z=0-1$ of L96 is attributed only to the lowest redshift data point of L96,
which is \cle 1.5$\sigma$ discrepant. While this is still consistent with our
measurement, the small difference is likely due to the much larger sample of
galaxies and the much larger surface area covered by the L96 sample as
compared to that of the relatively small HDF field.

\begin{table}[h]

\caption{Comoving UV Luminosity Density}
\begin{tabular}{lccccc}\tableline \tableline
Redshift & Luminosity Density\tablenotemark{a} & Luminosity Density & HDF Galaxies & HDF Galaxies \\
         &          & above SB Cut      &         & above SB Cut \\ \tableline
0.00--0.50 & 26.29$\pm$\err{0.31}{0.22} & 24.07$\pm$\err{0.24}{0.25} & 152 & 1 \\
0.50--1.00 & 26.43$\pm$\err{0.23}{0.14} & 24.98$\pm$\err{0.25}{0.26} & 241 & 7 \\
1.00--1.50 & 26.62$\pm$\err{0.24}{0.12} & 25.69$\pm$\err{0.11}{0.13} & 228 & 23 \\
1.50--2.00 & 26.65$\pm$\err{0.24}{0.12} & 25.61$\pm$\err{0.10}{0.10} & 205 & 49 \\
2.00--3.00 & 26.44$\pm$\err{0.28}{0.21} & 26.24$\pm$\err{0.17}{0.17} & 141 & 85 \\
3.00--4.00 & 26.41$\pm$\err{0.34}{0.27} & 26.15$\pm$\err{0.15}{0.14} & 62 & 60 \\
4.00--5.00 & 26.52$\pm$\err{0.44}{0.37} & 26.44$\pm$\err{0.19}{0.20} & 50 & 50 \\
5.00--6.00 & 26.16$\pm$\err{0.59}{0.38} & 25.77$\pm$\err{0.28}{0.28} & 8 & 8 \\ \tableline

\end{tabular}
\setlength{\baselineskip}{14.5pt}\vspace{14.5pt}  

\tablenotetext{a}{Values are log luminosity density at restframe 1500\AA\ in
units of $h_{100}^{-2}$ erg s$^{-1}$ Hz$^{-1}$ Mpc$^{-3}$ ($q_{\rm o}$=0.5).
Errors include the effects of systematic and photometric error on the
photometric redshifts and of sampling error.}

\end{table}

With sampling errors properly accounted for, the
errorbars generated from our bootstrap code are more than a factor of two
larger than those of C97 or M98 for the $z>2$ UV luminosity density. 
It can be seen that, after an increase at redshifts $z \simeq 0 - 2$,
the UV luminosity density remains constant to within errors to redshifts
$z\simeq 6$. In other words, {\em we find no convincing
evidence that the UV luminosity density decreases with redshift for $z>2$}.

\section{Effects of Cosmological $(1+z)^3$ Surface Brightness Dimming}

 To meaningfully compare the comoving UV luminosity density of the local,
low-redshift Universe with that of the distant, high-redshift Universe, it is
necessary to account for the very large effect of cosmological $(1+z)^3$
surface brightness (SB) dimming.  Because low-redshift galaxies are viewed to
much lower {\em intrinsic} SB thresholds than are high-redshift galaxies,
certain corrections must be applied in order to view all galaxies at a common
intrinsic SB threshold.

  We applied these corrections to our catalog of 1067 galaxies.
First, we corrected the flux of each galaxy to the value that would be
observed if the galaxy were placed at redshift $z = 5$.  Specifically, we
applied monochromatic SB corrections and $K$-corrections on a pixel-by-pixel
basis, as in Bouwens, Broadhurst, \& Silk (1998), and we also applied small
corrections to account for the different \wfpc\ passbands that sample
rest-frame 1500 \AA\ at different redshifts.

  Next, we applied a uniform SB cut on a pixel-by-pixel basis, excluding pixels
that failed to meet a minimum SB threshold.  The threshold was determined by
assuming that objects are detected in the F814W image to within $\sim 1 \sigma$
of sky, which corresponds to a SB of $1.27 \times 10^{-33}$ erg s$^{-1}$
cm$^{-2}$ Hz$^{-1}$ pixel$^{-1}$.  Next, we reversed the monochromatic SB
corrections and $K$-corrections on a pixel-by-pixel basis to bring each galaxy
back to its actual redshift and again calculated the comoving luminosity
density versus redshift.  In this way, only those parts of the galaxies that
are of high enough intrinsic SB to be detected at all redshifts up to $z =
5$ --- given the actual sensitivity of the HDF F814W image --- are included
into the measurement.

  Results are shown in the bottom panel of Figure 1, which plots the comoving
UV luminosity
density versus redshift of high intrinsic surface brightness regions, and
in the third column of Table 1. Also listed in Table 1 are the number of
galaxies within each redshift bin and the number of galaxies in each bin that
have at least one pixel above the SB cut. It is evident that (1) the
comoving UV luminosity density of high intrinsic surface brightness
regions increases by two orders of magnitude from $z\simeq 0$ to $z\simeq 5$,
and (2) star-forming objects
seen at $z > 2.5$ are relatively rare at $z < 2.5$. In other words, {\em the
comoving UV luminosity density contributed by high intrinsic SB regions
appears to increase monotonically with redshift, at least out to $z\simeq 5$}.

\section{Discussion and Conclusions}

  Previous determinations of the comoving UV luminosity density at $z>2$ rely
primarily on the M98 Lyman break galaxy sample. The stated uncertainties of
these measurements include contributions from incompleteness at the faint end
of the luminosity function (LF) as well as from the volume normalization and
color selection region.  While these effects do indeed contribute to the total
uncertainty, they are by no means the dominant factors.  For any LF with
power-law index $\alpha < 2$, the luminosity density is dominated
by the {\em bright} end of the LF.  Because the bright end of
the LF is inherently poorly sampled, this sampling error at
the bright end of the LF in fact dominates the uncertainty of
the luminosity density.  Indeed, at redshifts $z > 2$, our uncertainties
--- which include the effects of sampling error --- are more than two
times larger than those quoted by C97 and M98.  This calls into question the
statistical significant of the ``peak'' in the comoving UV luminosity density
at $z \simeq 1.5$.

Ignoring errors, one can see from Figure 1{\it a} that our $z=4$ measurement
of the UV luminosity density is higher than that of M98. It is possible that
this arises because the M98 method of finding galaxies at that redshift {\it
by their definition} does not find all galaxies at $z\simeq 4$. Rather, their
color-color polygon was designed to find objects that are almost certainly at
$z\simeq 4$ with little contamination from low-redshift objects, which
is why the data points of M96 were plotted as lower limits.
In Figure 2, we plot the
$z\simeq 4$ objects from our photometric redshift catalog with the M98 polygon.
One can see that even with the large ($B-V$) errors involved, the M98 technique
may be missing half of the high-redshift galaxies in the HDF.

\begin{figure}
\centerline{\psfig{file=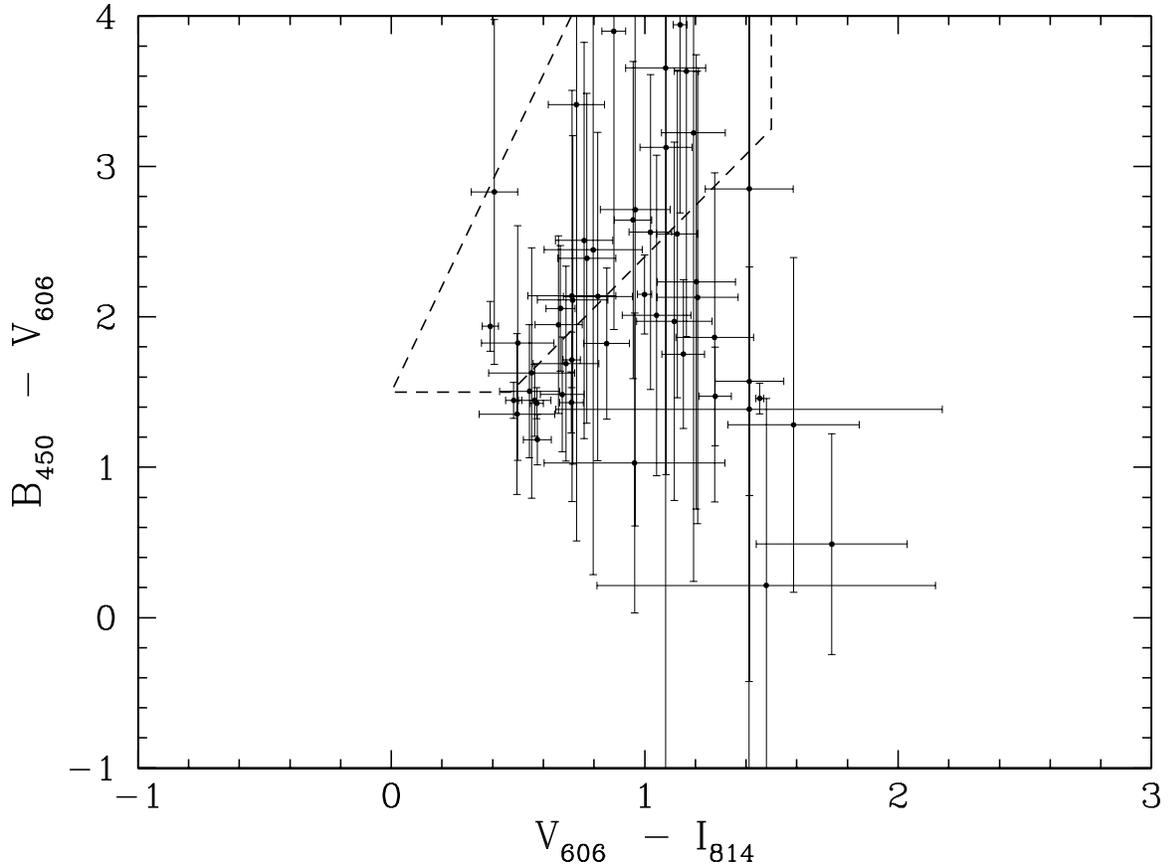,width=12cm}}
\caption{Color-color diagram similar to that of Madau \etal (1996, 1998)
containing $z\simeq 4$ galaxies selected from our photometric redshift catalog
of HDF galaxies and plotted with their measured photometric errors. It can be
seen that, while this technique does locate $z\simeq 4$ galaxies, almost 50\%
of them may lie outside the selection polygon despite the large ($B-V$) errors
involved.}
\end{figure}

Perhaps the single most important factor that must be included in such a
study is the enormous effect of the cosmological surface brightness dimming
at very high redshifts. At $z<0.1$ this effect is less than a factor of 1.3,
but at $z=5$
it becomes a factor of 216. Therefore, if we are to compare the UV luminosity
density at $z=5$ to that of the local Universe, we must consider local
galaxies only down to the same SB level as that sampled at high-redshift.

At the highest redshifts, the severe cosmological SB dimming affecting the HDF
observations allows only the very highest levels of UV luminosity ``column
density'' (or equivalently, star formation rate ``column density'') to be
sampled. At progressively lower redshifts, we are able to observe galaxies
down to lower and lower SB thresholds. The uniform SB cut that we applied to
our galaxy sample to allow for this discrepancy corresponds to a star formation
rate column density of $\simeq 0.1 h_{100}^2$ M$_{\sun}$
yr$^{-1}$ kpc$^{-2}$ at $z=5$ (assuming $q_{\rm o}=0.5$ and using the relation
of UV luminosity to star formation rate with a Salpeter IMF from M98). As
seen in Table 1 and the bottom panel of Figure 1, almost no $z<1$ objects have
star formation rates above this level, while at high redshifts, there are many
objects exceeding this cut. For example, only 7 out of 241 galaxies from the
$z=0.5-1.0$ redshift bin would be visible at $z=5$, and then only the
brightest few image pixels of those objects would peak above the SB cutoff.

Figure 1 provides us with another piece of evidence in favor of a UV
luminosity density that increases with redshift. In the lowest redshift bin
($z=0.0-0.5$) of Figure 1{\it a} and {\it b}, where the SB effect is smallest,
the intrinsically high SB regions make up only 0.5\% of the total UV
luminosity density. If this ratio is the same at high redshifts as it is
at low redshifts then our measurements of the UV luminosity density at high
redshifts in Figure 1{\it a} may need to be increased by up to a factor of
150. Although this ratio may be quite different at high redshifts than it is at
low redshifts, this suggests that the UV luminosity density could be a strongly
increasing function of redshift.

We have shown that when the low- and high-redshift Universes are observed on
equal footing, one gets a completely different impression of the global UV
luminosity density than previously thought. We find that objects with star
formation rates comparable to those at $z$\cge 3 are very rare in the nearby
Universe. This implies that a majority of the star formation may have occurred
at very high redshifts, and therefore that a peak in the
star formation rate density of the Universe has not yet been observed and
likely lies somewhere at $z>2$.

\acknowledgments

SMP and KML acknowledge support from NASA grant NAGW-4422 and NSF grant
AST-9624216. AF acknowledges support from a grant from the Australian
Research Council.

\end{document}